\documentclass[2opt]{article}
\usepackage{amsmath, amssymb, booktabs}
\usepackage{amsthm, multirow}
\usepackage{mathtools}
\usepackage{booktabs}
\usepackage{enumerate}
\usepackage{upquote}
\usepackage[margin = 1in]{geometry}
\usepackage{graphicx}
\usepackage{setspace}
\usepackage{subcaption}
\usepackage{tikz}
\usetikzlibrary{positioning, calc, shapes.geometric, shapes.multipart,
                shapes, arrows.meta, arrows,
                decorations.markings, external, trees}
\tikzstyle{Arrow} = [
    thick,
    decoration={
        markings,
        mark=at position 1 with {
            \arrow[thick]{latex}
            }
        },
    shorten >= 3pt, preaction = {decorate}
    ]           
\usepackage[utf8]{inputenc}
\usepackage[font=footnotesize,labelfont=bf]{caption}
\newcommand{\indep}{\rotatebox[origin=c]{90}{$\models$}}
\title{Regression to the Mean's Impact on the Synthetic Control Method: Bias and Sensitivity Analysis}
\author{Nicholas A. Illenberger, Dylan S. Small, Pamela A. Shaw}
\date{}

\begin{document}
\maketitle

\begin{abstract}
    To make informed policy recommendations from observational data, we must be able to discern true treatment effects from random noise and effects due to confounding. Difference-in-Difference techniques which match treated units to control units based on pre-treatment outcomes, such as the synthetic control approach, have been presented as principled methods to account for confounding. However, we show that use of synthetic controls or other matching procedures can introduce regression to the mean (RTM) bias into estimates of the average treatment effect on the treated. Through simulations, we show RTM bias can lead to inflated type I error rates as well as decreased power in typical policy evaluation settings. Further, we provide a novel correction for RTM bias which can reduce bias and attain appropriate type I error rates. This correction can be used to perform a sensitivity analysis which determines how results may be affected by RTM. We use our proposed correction and sensitivity analysis to reanalyze data concerning the effects of California's Proposition 99, a large-scale tobacco control program, on statewide smoking rates.
\end{abstract}

\textbf{Keywords}: Regression to the mean, difference-in-difference, matching, synthetic controls

\section{Introduction} \label{section:intro}

\par In observational research we must grapple with the fact that observed differences between pre-treatment and post-treatment outcomes may not be the result of treatment alone. Outside events, random error, and outcome trends unrelated to treatment can add noise to our data, making it difficult to determine the effects of an intervention. The difference-in-difference (DID) estimator \cite{angrist2008mostly} attempts to provide a solution to this problem. 
\par Given a treated and control unit with outcomes measured pre- and post-intervention, DID is the difference in pre-treatment outcomes subtracted from the difference in post-treatment outcomes. If we assume that the treated and control groups would have parallel outcome trends in the absence of treatment, then the estimator is unbiased for the average treatment effect on the treated (ATT) even in the presence of unmeasured confounding. Ease of use and robustness to unmeasured confounding, has made the DID popular among epidemiologists. The effects of background checks on firearm homicide and suicide rates \cite{kagawa2018repeal}, the development of urban green space on crime rates \cite{branas2011difference}, and policy changes on the health of mothers receiving nutritional support \cite{hamad2019impact} have all been assessed using the DID framework. However, because the estimator is not robust to deviations from the parallel trends assumption, control units must be selected with care.
\par To improve the selection of controls, Abadie et al. \cite{abadie2005semiparametric} introduced the method of synthetic controls. Here, a new control unit is constructed using a weighted sum of donor controls. If we weigh the donors such that they closely resemble the treated unit in the pre-intervention period, then under standard assumptions they should be a plausible representation of how the treated unit would look without treatment in the post-intervention period. This method works similarly to matching. Control units which are more similar to the treated unit are given larger weights in the synthetic control than those which are dissimilar. Because the synthetic control method uses information from all control units, estimates of the ATT will tend to be less variable than estimates obtained from one-to-one matching techniques. While matching may improve comparability between treated and control units, recent work by Daw et al.\cite{daw2018matching} have shown that nearest-neighbor matching can induce bias in DID estimates due to regression to the mean (RTM). Because of the similarities between matching and the synthetic control method, there is a need to better understand how RTM can affect the synthetic control estimator.

\par In this paper, we examine the effect of RTM on estimates of the ATT coming from the synthetic control and other matched DID methods. Through simulations, we show that RTM can result in inflated type I error rates and, in some settings, decreased power. Compared to other matching techniques, these effects are exaggerated in the synthetic control estimator. We also propose a novel sensitivity analysis, which can be used to check how robust inference may be to the effect of RTM bias. Sensitivity and quantitative bias analyses allow researchers to assess the potential effects of systematic error in an experiment \cite{lash2014good}. These approaches are common in causal inference and missing data settings \cite{robins2000sensitivity}, and have been used to quantify the uncertainty associated with measurement error \cite{tordoff2019misclassification}. We apply our proposed sensitivity analysis to reanalyze data from Abadie et al.\cite{abadie2010synthetic}, estimating the effect of a large-scale tobacco control initiative on smoking cessation rates in California.

\section{Methods} \label{section:methods}

\subsection{Matched Difference-in-Difference} \label{sec:mdid}

\par Consider a scenario in which we have two time periods, $t=0$ for the pre-treatment period and $t = 1$ for the post-treatment period. Let $Y(t)$ represent the observed outcome at time $t$, $A$ be an indicator of treatment status, and let $X$ be either measured or unmeasured confounders. We further define $Y^a(t)$ to be the potential outcome \cite{rubin2005causal} which would be observed under treatment $A=a$ at time $t$. In this setting $Y^1(0) = Y^0(0)$ because neither group receives treatment at time $t=0$. The DID estimator assumes the linear model, $\mathbb{E}\left[Y^0(t)|X \right] = \beta X + \gamma t$, for the expected potential outcome under no treatment at time $t$. The distribution of $X$ will typically differ between the group truly receiving treatment and the control group. Because of this, we expect that the potential mean under no treatment will differ between the treatment groups.

\par Note that the effect of time, $\gamma$, does not depend on confounders. Likewise the effect of confounders, $\beta$, does not depend on time. These are jointly known as the parallel trends assumption. If both are true and if the distribution of covariates within each group remains the same from period to period, then the expected difference between the potential untreated outcomes for the the treated and control units in the pre-treatment period is equivalent to that in the post-treatment period. This is the case regardless of the value of confounders, $X$, within these units. We define $D(t)$ to be the expected difference between the treatment and control groups at time $t$. Alongside these homogeneity assumptions, we also assume consistency, $Y^A(t) = Y(t)$, and $Y^0(t)\indep A|X$. Consistency states that the potential outcome under the treatment received is equivalent to the observed outcome. The second assumption, that the potential untreated outcome is independent of treatment conditional on $X$, is likely to hold because $X$ can contain any measured or unmeasured confounders. Under these assumptions, we can show (see Appendix):
\begin{align*}
    D(t) &= \mathbb{E}\left[Y(t)|A=1 \right] - \mathbb{E}\left[Y(t)|A=0 \right] \\
    &= \mathbb{E}\left[ Y^1(t)-Y^0(t)|A=1\right] + \beta\left(\mathbb{E}[X|A=1] - \mathbb{E}[X|A=0] \right).
\end{align*}

\par For $t=1$, the first term in this summation is the ATT, which we define as $\theta$. Because $Y^1(0) = Y^0(0)$, it follows that $\theta$ is the difference between $D(1)$ and $D(0)$. The DID estimator suggests using the empirical means within treatment groups at times $t=0$ and $t=1$ to estimate this difference. That is, $\hat{\theta} = \widehat{D}(1) - \widehat{D}(0)$, where
 
\begin{equation*}
    \widehat{D}(t) = \frac{\sum_{i=1}^n Y_{i}(t)\mathbb{I}(A_i = 1)}{\sum_{j=1}^n \mathbb{I}(A_j = 1)} - \frac{\sum_{i=1}^n Y_{i}(t)\mathbb{I}(A_i = 0)}{\sum_{j=1}^n \mathbb{I}(A_j = 0)}.
\end{equation*}

\par In practice, it may be difficult to identify units such that $\beta$ and $\gamma$ are equivalent in the treated and control groups. Ryan et al. \cite{ryan2015we} have shown that matching can decrease bias in some cases by improving the comparability of units. However, as we will discuss, matching can have the effect of introducing bias into estimates of the DID.

\subsection{Regression to the Mean} \label{sec:rtm}

\par Regression to the mean (RTM) is a statistical phenomena in which extreme measurements of a random variable tend towards their expected value upon repeat measurement. If not properly accounted for, RTM can lead to misleading results. Compared with the widespread use of matched DID methods, relatively little work has been done to examine the effects of RTM in this setting. Daw et al. \cite{daw2018matching} have found that nearest-neighbor matching can induce bias in estimates of the ATT when using DID techniques. However, as far as we are aware, no work has examined the effects of RTM on the synthetic control method nor on the statistical power of inference based on matched DID methods.
\par To build intuition for how matched DID estimators are subject to RTM bias, consider the following example. We are interested in how a proposed treatment affects the trajectory of an outcome variable. A single unit which will receive treatment and a number of control units are sampled. Bias due to RTM can be introduced when there is variability in outcome measures and the population from which the treated unit is drawn differs from the control population. Suppose the pre-treatment outcome measurements for the control and treatment populations are normally distributed with mean $\mu_0$ and $\mu_1$ respectively. Without loss of generality, assume $\mu_1 > \mu_0$. The nearest-neighbor match for the treated unit is expected to be a control unit with an observed pre-treatment measurement greater than its expected value. Thus, even if the post-treatment outcome distributions for the treated and control populations are equivalent to those of the pre-treatment period (i.e. no treatment effect), the matched unit is expected to decrease upon repeat measurement. This creates the false appearance of a treatment effect. Specifically, because the outcome trajectory of the matched control is expected to be decreasing, the parallel trends assumption is violated and estimates of the ATT are invalid.

\subsection{Matching procedures}\label{sec:matching}

\subsubsection{Synthetic Control Method}
The method of synthetic controls is provided in detail elsewhere \cite{abadie2010synthetic}, and so we provide only a brief overview. Suppose we collect data on a single treated unit and $n_0$ controls for a total of $n_0 + 1$ units. Without loss of generality, let $i = 1$ denote the treated unit and $\mathcal{C}$ denote the set of indices for the control units. We collect $\tau$ outcome measurements $Y_i = (Y_{i1}, \hdots Y_{i\tau})$ on each unit. Suppose that treatment is withheld until time $\tau_0$, such that $j \in \{ 1, \hdots, \tau_0\}$ denote the pre-treatment period and $j \in \{ \tau_0 + 1, \hdots, \tau \}$ compose the post-treatment period. The synthetic control method entails finding a set of non-negative weights such that, at each pre-treatment timepoint, the weighted sum of the control units' outcomes closely resembles those of the treated unit. That is, we aim to select $w_k$, for $k \in \mathcal{C}$ such that $Y_{1j} \approx \sum_{k\in\mathcal{C}} w_k Y_{kj}$ for $j \in \{1, \ldots, \tau_0\}$ and $\sum_{k\in\mathcal{C}} w_k = 1$. If weights are chosen so that these equalities approximately hold, then the weighted sum of the post-treatment control vectors can serve as a potential untreated outcome vector for the treated unit. 

\subsubsection{Nearest Neighbor matching}
\par Nearest neighbor matching is a procedure which matches the treated unit with the single control unit which is ``closest" to it in the pre-treatment period. To be concrete, let $S$ be a set of pre-treatment outcome measurements observed in a sample. Suppose we have another unit with pre-treatment vector $s$ and we want to find the nearest neighbor match for $s$ over the set $S$. The nearest neighbor match given a selected distance metric is the element of $S$ which minimizes the distance from $s$ \cite{arya1994optimal}. Note that different distance metrics may result in different matches. In this paper, we consider two implementations of nearest neighbor matching. The first method is based upon the distance between pre-treatment outcome vectors as determined by the $L_2$-norm, while the second uses the $L_1$ distance between coefficients in an OLS regression of pre-treatment outcome measurements on time (i.e. pre-treatment trend).

\section{Simulations}\label{sec:simulations}

\par To examine the effect of RTM bias, we simulate a single treated unit alongside $n_0 = 40$ controls. For control units, eight outcome measurements are drawn from a multivariate normal distribution with mean $\mu_0$, marginal variance $\sigma^2 = 1$, and first order autoregressive (AR(1)) covariance structure with correlation $\rho^{|t_i - t_j|}$ between outcome measurements at $t_i$ and $t_j$. The treatment unit is simulated similarly, with mean $\mu_1$ rather than $\mu_0$. For each simulated dataset we match our treated unit to controls using the synthetic control method, nearest neighbor based on the $L_2$-norm, and nearest neighbor based on pre-treatment trend. For comparison, we provide an estimate of the treatment effect using the unmatched DID. If we define $\bar{Y}_{0j} = n_0^{-1}\sum_{k\in \mathcal{C}} Y_{kj}$ as the mean of the control units' outcomes at time $j$, then the unmatched DID is calculated as

\begin{equation*}
   \hat{\theta} = \frac{1}{4}\left[\sum_{j=5}^8 \left\{ Y_{1j} - \bar{Y}_{0j}\right\} - \sum_{j=1}^4 \left\{ Y_{1j} - \bar{Y}_{0j}\right\}\right]
\end{equation*}

For the nearest neighbor and synthetic control methods, the estimator for treatment effect simply replaces $\bar{Y}_{0j}$ with the value of the matched or synthetic control at time $j$.

\subsection{Type I errors and bias}

\par Our first set of simulations looked to determine how type I error rates are affected by outcome level matching. Permutation tests were used to test the null hypothesis of no treatment effect. For each unit $i = 1,\hdots, n_0+1$, we relabel the data so that individual $i$ is the treated unit and all other units are in the control group. We estimate the treatment effect under each of the relabellings to obtain $\hat{\theta}_i$ for $i = 1,\hdots, n_0+1$. The p-value for this test is given as $p = n^{-1}\sum \mathbb{I}(|\hat{\theta}_1| \ge |\hat{\theta}_i|)$. For sufficiently low p-value, we reject the null hypothesis of no treatment effect. This test is equivalent to the "placebo test" described in Abadie et al. \cite{abadie2010synthetic}.
\par First, we fix $\mu_0 = 0$ and $\rho = 0.5$, while varying the value of $\mu_{1}$. The magnitude of bias induced by regression to the mean is directly related to the difference between the expected value of the controls' outcomes and those of the treated. Because of increased bias, type I error rates should increase as the distance between $\mu_0$ and $\mu_1$ grows. This effect is seen in the left-hand side of Table \ref{tab:mu1}. Type I error rates were based on $2000$ simulations where we reject the null hypothesis for p-value lower than $0.05$. On the right-hand side, we provide the results from varying the value of $\rho$, which controls the correlation between repeat measurements. For these simulations $\mu_1$ is set to $5$. As the correlation increases, we observe that the type I error rate decreases. Errors which are highly correlated are less prone to large shifts towards the mean, resulting in less bias in the short term. In both scenarios, the synthetic controls method exaggerates the effects of RTM bias when compared to the nearest neighbor methods. Again, because the synthetic control uses information from all control units, there is less variance in the estimator. Using synthetic controls results in more confidence in the biased estimate. In both scenarios, matching based on pre-treatment linear trend did not increase type I error rates. This is consistent with findings from Daw et al. \cite{daw2018matching} showing that this matching did not introduce perceptible bias into estimates of the ATT.

\begin{table}[t]
    \centering
    \begin{tabular}{c cccc cc cccc}
    \toprule
    \multicolumn{5}{c}{\underline{Type I Error Rate: Varying $\mu_1$}} & & \multicolumn{5}{c}{\underline{Type I Error Rate: Varying $\rho$}} \\
    $\mu_{1}$ & Unmatched & SC & $NN_1$ & $NN_2$ & & $\rho$ & Unmatched & SC & $NN_1$ & $NN_2$ \\
    \midrule
    1.00 &  0.05 & 0.16 & 0.09 & 0.05 & &0.00 &  0.05 & 0.40 & 0.29 & 0.06\\
    2.00 &  0.05 & 0.31 & 0.19 & 0.05 & &0.25 &  0.04 & 0.39 & 0.29 & 0.05\\
    3.00 &  0.05 & 0.35 & 0.25 & 0.05 & &0.50 &  0.04 & 0.36 & 0.27 & 0.05\\
    4.00 &  0.05 & 0.35 & 0.26 & 0.05 & &0.75 &  0.05 & 0.25 & 0.18 & 0.05\\
    5.00 &  0.04 & 0.33 & 0.25 & 0.05 & &0.90 &  0.05 & 0.18 & 0.13 & 0.05 \\
    \midrule
    \end{tabular}
    \caption{Type I error rates for the unmatched DID, synthetic control (SC), nearest neighbor using the $L_2$-norm $(NN_1)$, and nearest neighbor using linear trends $(NN_2)$.}
    \label{tab:mu1}
\end{table}

\subsection{Loss of power}

\par Matching on pre-treatment outcomes can cause us to reject the null hypothesis more often than desired when there is no treatment effect. However, when there is a treatment effect, RTM can also result in decreased power.  Imagine a scenario where $\mu_1$ is greater than $\mu_0$. As we have seen, matched control units are likely to be those with randomly high pre-treatment outcome levels. If there is no treatment effect, then the post-treatment difference between the treated and matched control unit is expected to increase because the control unit's outcome measurements are expected to decrease. However, if there is a treatment effect which causes the treated unit's outcome level to decrease, then the difference in post-treatment measurements will likely not be as large as when there is no treatment effect. Thus, when the treatment effect is in the same direction as RTM, matching on outcome levels can make us less likely to reject the null hypothesis when there is a treatment effect and more likely when there is not. Likewise, when the treatment effect is in the opposite direction of RTM bias, the effect of treatment is expected to be exaggerated. 
\par
\begin{figure}
    \centering
    \includegraphics[scale = 0.3]{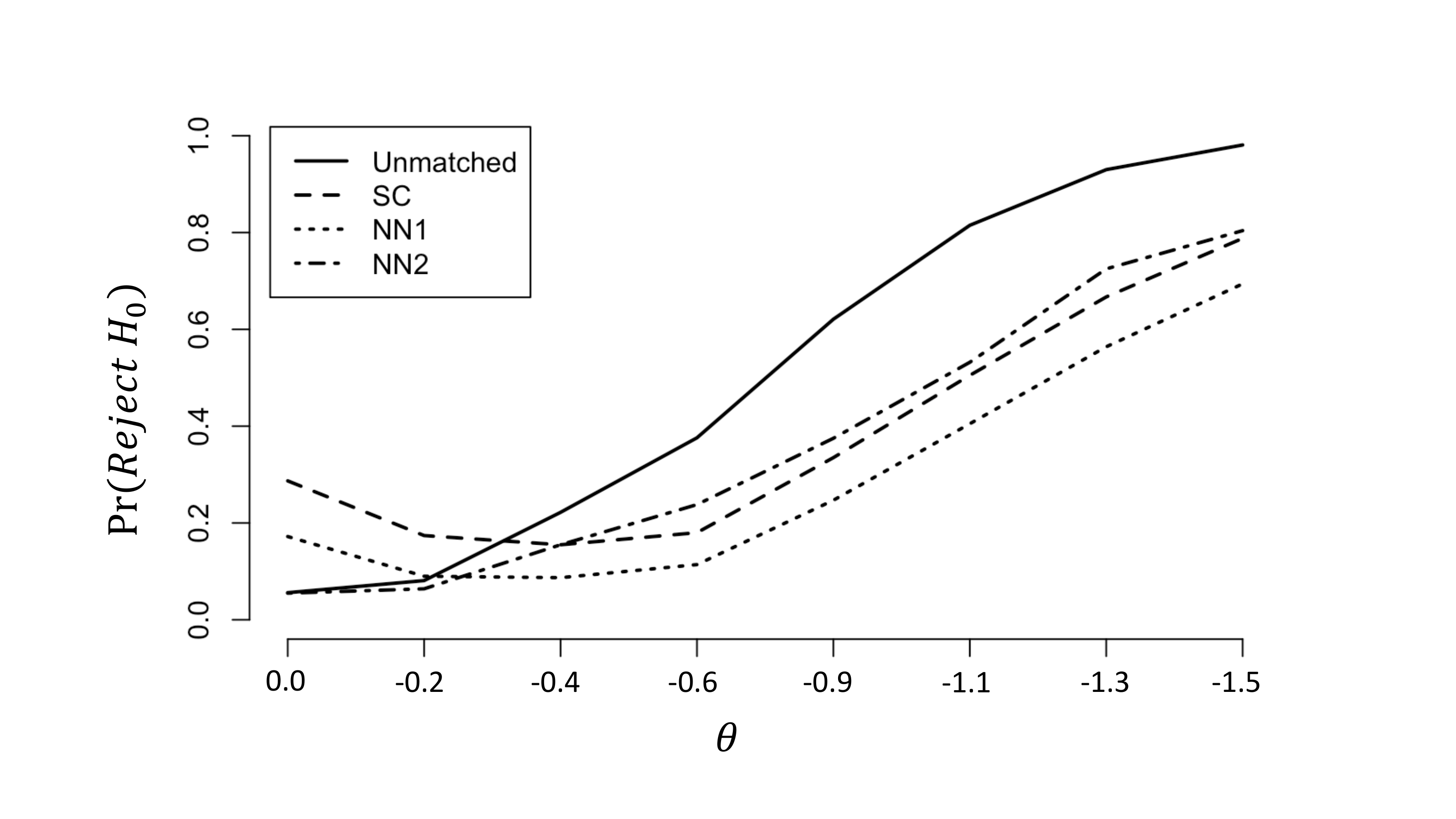}
    \caption{Empirical probability of rejecting the hypothesis that there is no treatment effect $(\theta = 0)$ as a function of $\theta$ using the unmatched, synthetic control method (SC), nearest neighbor matching on $L_2$-norm (NN1), and nearest neighbor matching on linear trend (NN2).}
    \label{fig:power_curve}
\end{figure}

 To illustrate this phenomena, we perform $2000$ simulations with $\mu_0 = 0$, $\mu_1 = 5$, and $\rho = 0.5$. Here, we induce a treatment effect, $\theta$. For each additional time point in the treatment period, the expected outcome for the treated unit increased by $\theta$. Note that for negative $\theta$, the treatment effect and RTM are working in the same direction and decreased power is a concern. Figure \ref{fig:power_curve} provides rejection rates for the unmatched and synthetic control procedure when $\theta$ is between $0$ and $-1.5$. As in the previous simulations, we see that when there is no treatment effect, the synthetic control method exhibits inflated type I error rates while the unmatched data has appropriate rejection rates. As the treatment effect increases, the rejection rate of the unmatched estimator's power surpasses that of the synthetic control method. These results indicate that depending on the direction of the treatment effect in relation to the direction of RTM, the synthetic control method can result in either conservative or anti-conservative bias.

\subsection{Correction and Sensitivity Analysis}

\par Suppose $Y_1, \hdots, Y_{T}$ are jointly normal random variables. By properties of the multivariate normal distribution, for any given $i$ and $j$, we have $\mathbb{E}\left[ Y_i | Y_j \right] = \mu_i + \Sigma_{ij} \Sigma_{jj}^{-1}(Y_j - \mu_j)$. If we know the mean at each time point and the covariance structure, then we can use this representation to account for RTM bias in matched DID estimates. To illustrate, suppose we are performing a DID analysis using a single treated observation and a sample of control units. Using a one-to-one matching technique, the treated unit $Y_1$ is matched with a control unit, $Y_m$, based upon pre-treatment outcome levels to obtain an estimate of the ATT, say $\hat{\theta}_{obs}$. This estimate can be conceptualized as the sum of the effect due to RTM bias and the effect due to treatment. Our correction technique subtracts the estimated effect of RTM, $\hat{\theta}_{rtm}$, from the observed effect to obtain a bias-adjusted estimate of the ATT, $\hat{\theta}_{adj} = \hat{\theta}_{obs} - \hat{\theta}_{rtm}$.

\par To obtain $\hat{\theta}_{rtm}$, define $\hat{Y}_{ij} = \mu_{ij} + \Sigma_{j\tau_0} \Sigma{\tau_0 \tau_0}^{-1} \left(Y_{i\tau_0} - \mu_{1\tau_0}\right)$ for $j > \tau_0$ and $i \in \{1, m\}$. Here, $\tau_0$ is the final pre-treatment observation time and $\mu_{ij} = \mathbb{E}[Y^0_i(j)]$ is the expected potential outcome level under no treatment for unit $i$ at post-treatment time $j$. Thus, $\hat{Y}_{ij}$ is the expected observation for unit $i$ at post-treatment time $j$ conditional on the final pre-treatment observation assuming no treatment effect. These expected values incorporate the effect of RTM. Calculating the DID using these expected values in place of observed post-treatment values for the treated and matched control units provides $\hat{\theta}_{rtm}$, which can be used to find $\hat{\theta}_{adj}$. Note that for one-to-one matching we only need to calculate $\hat{Y}_{ij}$ for the treated and matched units. For the synthetic control correction, we will need to estimate this value for all units.
\par To generalize this adjustment for use with synthetic controls, first obtain the synthetic control weights $w_k$ for $k \in \mathcal{C}$. Using these weights, we construct a synthetic outcome vector $Y_S$, where $Y_{Sj} = \sum_{k \in \mathcal{C}} w_k Y_{kj}$, and find the observed estimate of the ATT, $\hat{\theta}_{obs}$. We can obtain the expected DID under RTM by using an augmented synthetic control. This new synthetic control is constructed using the same weights as before and replacing post-treatment control measurements with the $\hat{Y}_{ij}$'s defined earlier. Call this augmented control $\hat{Y}_{S}$ and calculate the expected DID under no treatment effect, $\hat{\theta}_{rtm}$, by subtracting the mean difference in observed pre-treatment outcomes of the treated unit and the synthetic control unit from the mean difference in expected post-treatment outcomes.
\par As a proof of concept, we perform $2000$ simulations with outcomes drawn from a multivariate normal distribution with AR(1) error structure. Here, $\mu_0 = 0$, $\mu_1 = 1$, $\sigma^2 = 1$, $\rho = 0.5$, and there is no treatment effect. For each simulation, we test the null hypothesis of no treatment effect using the permutation test described earlier, replacing $\hat{\theta}_i$ with $\hat{\theta}_{i,adj}$. This correction was derived under the assumption of multivariate normality. To test if it is robust to deviations from this assumption, we also look for type I error rates when outcomes are drawn from a multivariate $t$-distribution. Simulation results are given in Table \ref{tab:terrors}. When errors are normally distributed, the adjusted synthetic control estimate of the ATT attains nominal type I error rates. We note that error rates are inflated for $t$-distributed outcomes, particularly for high levels of $\rho$. However, observed error rates are lower than those obtained in Table \ref{tab:mu1} using the unadjusted synthetic control approach with normally distributed errors.

\begin{table}[t]
    \centering
    \begin{tabular}{cccc}
    \toprule
        Degrees of Freedom &  $\rho = 0.25$ & $\rho = 0.50$ &  $\rho = 0.75$\\
    \midrule
    \midrule
        $\infty$ ($\text{Normal}$) & 0.05 & 0.05 & 0.05 \\
        50     & 0.05  & 0.05 & 0.06 \\
        10     & 0.05  & 0.06 & 0.08 \\
        3      & 0.05  & 0.08 & 0.12 \\
    \bottomrule
    \end{tabular}
    \caption{Type I error rates for the adjusted DID estimator when normality assumption is not satisfied. Errors come from a t-distribution with degrees of freedom described.}
    \label{tab:terrors}
\end{table}

\par In practice, we do not have access to the true values of $\mu_{1j}$, $\mu_{0j}$, $\rho$, or $\sigma$. So, estimating $\hat{\theta}_{adj}$ is not possible without making additional assumptions on the values of these parameters. To address this problem, we propose treating these as sensitivity parameters. By positing a range of values for these parameters and calculating $\hat{\theta}_{adj}$ under each set, we can quantify how much our estimate of the ATT may be affected by RTM. For example, suppose we have obtained an estimate of the ATT from a matched DID analysis and have determined that the effect is significant using a permutation test. We wish to determine if the observed significance can be explained away by RTM. To do so, we select values of the sensitivity parameters under which there is no treatment effect, estimate $\hat{\theta}_{adj}$ using these, and then perform the permutation test. If the sensitivity parameters must be extreme or implausible in order to nullify significance, then our results are unlikely to have been the result of RTM bias.

\section{Reanalysis of smoking cessation data}\label{sec:analysis}

\par To further understanding of our proposed sensitivity analysis, we reanalyze data from Abadie et al. \cite{abadie2010synthetic} concerning the effect of California's Proposition 99 on smoking cessation. Proposition 99 was an anti-smoking initiative ratified by California state voters in 1988. The act added a 25 cents per pack tax on the sale of cigarettes, and earmarked tax revenue for use in health care programs and anti-tobacco advertisements.
\par The purpose of the original analysis was to determine whether Proposition 99 was successful in decreasing tobacco consumption. The authors concluded that the initiative decreased cigarette consumption in California by about 20 packs per capita annually. This difference was found to be statistically significant using a permutation test. However, their analysis was based upon the synthetic control method, which we have shown induces bias in estimates of the DID. To determine the robustness of the original findings, we will perform a sensitivity analysis.

\begin{figure}
    \centering
    \begin{minipage}{0.6\textwidth}
    \centering
    \centering
    \begin{tabular}{l cc}
    \toprule
    Variable & Effect size & Standard Error \\
    \midrule
    \midrule
    Year & -1.28 & 0.16  \\
    log(GDP) & -5.85 & 6.54  \\
    Beer consumption & -0.03 & 0.10  \\
    $\%$ Age 15 to 24 & -118.97 & 100.00  \\
    Retail price of cigarettes & 0.02 & 0.012  \\
    Cigarette sales per capita (1975) & 0.11 & 0.08 \\
    Cigarette sales per capita (1980) & 0.44 & 0.17 \\
    Cigarette sales per capita (1988) & 0.38 & 0.11 \\
    \bottomrule
    \end{tabular}
    \end{minipage}\hspace{0.5cm}
    \begin{minipage}{0.3\textwidth}
    \centering
    \includegraphics[scale = 0.4]{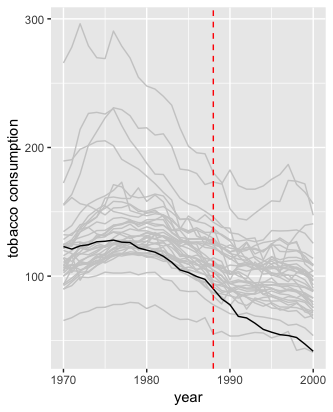}
    \end{minipage}
    \caption{Left: Coefficients from fit GEE with AR(1) working correlation matrix. Right: Tobacco consumption (per capita cigarette consumption) in a subset of states between 1970-2000. California highlighted in black, treatment initiation indicated by dashed red line.}
    \label{fig:subset_reanalysis}
\end{figure}

\par This analysis was based upon cigarette consumption rates in 39 states, California and 38 controls. Eleven states which implemented large-scale tobacco control programs or that raised taxes on cigarette sales were excluded from the analysis. The right-hand side of Figure \ref{fig:subset_reanalysis} provides a plot of cigarette consumption rates between 1970 and 2000 for the included states are provided. 
\par As in Abadie et al., we use logged per capita GDP, the average retail price of cigarettes within each state, beer consumption per capita, and the percentage of the population aged 15 to 24 to construct our synthetic control unit. Following the example of the original study we also included cigarette sales in the pre-treatment years 1975, 1980, and 1988. These variables were chosen specifically because they were correlated with outcome levels. However, this has a similar effect to matching on outcome levels, making RTM bias a concern. Without adjusting for induced RTM bias, we ended up with an estimated ATT of $21.25$ and a p-value of $2/39 \approx 0.05$. That is, in the period from 1989 to 2000, after Proposition 99 had been enacted, the unadjusted estimate determined that Californians consumed about 20 fewer packs of cigarettes per capita than they would have if the law had not taken effect.
\par Our proposed method for adjusting the DID relies on knowledge of the mean model under no treatment effect for each group. To obtain plausible mean models, we regress per capita cigarette sales on the variables previously used in the synthetic control method. Using the generalized estimating equation (GEE) framework with AR(1) working correlation matrix, we obtain linear mean models for each state's cigarette consumption. In the left-hand side of Figure \ref{fig:subset_reanalysis}, estimated effect sizes and standard errors corresponding to the different predictors are provided. We see that there is a strong downwards trajectory in cigarette consumption over time in our dataset. To determine the error structure (i.e. variance and correlation parameters) for our correction, we calculate the sample residual variance, $\hat{s^2}$, and the sample correlation, $\hat{\rho}$, between adjacent residuals within each state. Respectively these values were estimated as $14.6$ and $0.93$.
\par Next, we use these values to calculate the expected DID due to RTM. If $g_i(j)$ is the fitted value from the GEE model for state $i$ at time $j$, then we define $\hat{Y}_{ij} =  \hat{\rho}^{j - 1988} \left(Y_{i,1988} - g_1(1988)\right) + g_i(j)$ for $j > 1988$ and $i \in \{1, \hdots, 39\}$. Using the estimated $\hat{Y}_i$ vectors, we obtain an RTM adjusted estimate of the ATT $20.76$ and an updated $p$-value of $4/39$ or $0.10$. While adjustment has a small effect on our estimate of the DID, our results are no longer significant at the $0.05$ level.
\par To illustrate how our sensitivity analysis would work, suppose we are interested in significance at the $0.1$ level. Consider the set of potential mean models for the treated unit defined by $g^*_1(j) = g_1(j) + \Delta$. For the purposes of this illustration, imagine that when $\Delta = 0$ as in the correction above, the test is still significant at the $0.05$ level. If small changes in $\Delta$ result in flipped significance, then the results are not robust to RTM. We re-estimate $\hat{\theta}_{adj}$ for a range of $\Delta$. When $\Delta = -1$, our adjustment leads to an estimate of $20.41$ and a p-value of $4/39$ which is greater than $0.1$. Note that $\Delta$ shifts the mean model for per capita cigarette consumption. Comparing with the scale of Figure \ref{fig:subset_reanalysis}, we see that this is a relatively minor shift. Because the sensitivity parameter, $\Delta$, did not need to be large to cause our results to be insignificant at the $0.1$ level, we have evidence that the results of the original study are not robust to RTM bias.

\section{Discussion}

\par In this paper we have illustrated the effects of RTM bias on the matched DID estimator. This extends upon work done by Daw et al. \cite{daw2018matching} showing the bias induced by nearest neighbor matching. Here, we have shown how this bias also occurs when using the method of synthetic controls and have provided simulations showing the effect of this bias on type I error rates and power. Our results suggest that practitioners should be more cautious of RTM bias when using synthetic controls than when using nearest neighbor matching. The added "confidence" in the model, gained from utilizing information from all of the control units, increased the type I error rate by a factor of two over the nearest neighbor approach.
\par We have further built on this work by developing an approach to determine how sensitive matched DID estimates may be to RTM bias. Sensitivity analyses are recognized as a critical component of causal analyses which allow us to determine how robust our results may be to deviations from assumptions. Using our approach, we showed that results concerning the effect of Proposition 99 on cigarette consumption in California may be overstated. After adjusting for plausible bias due to RTM, the results of the analysis in Abadie et al. \cite{abadie2010synthetic} were no longer significant.
\par In the future, it may be worthwhile to look for ways to correct for RTM bias when we can not assume normality of errors. For $t$-distributed errors, we noticed that type I error rates were slightly greater than desired $\alpha$-levels. While the adjustment still performed better than the unadjusted synthetic control estimator, we believe the method could be improved upon. As a whole, we believe that when researchers apply matched DID estimators, they should also provide evidence that their results are robust to RTM bias, either by using our adjusted DID estimator or by providing a sensitivity analysis.

\nocite{*}
\bibliographystyle{plain}

\newpage

\appendix
\section{Appendix}
We wish to prove:
\begin{equation*}
    D(t) = \mathbb{E}[Y^1(t) - Y^0(t)|A=1] - \beta(\mathbb{E}[X|A=1] - \mathbb{E}[Y|A=0])
\end{equation*}
\begin{proof}
Consider the following:
\begin{align*}
    D(t) &= \mathbb{E}[Y(t)|A=1] - \mathbb{E}[Y(t)|A=0] \\
         &= \mathbb{E}[Y^1(t)|A=1] - \mathbb{E}[Y^0(t)|A=0] \\
         &= \mathbb{E}[Y^1(t) - Y^0(t)|A=1] + \mathbb{E}[Y^0(t)|A=1] - \mathbb{E}[Y^0(t)|A=0]
\end{align*}
Here, the second line follows from the consistency assumption. Next note, for $A = 0$ or $1$, the expected value of the potential distribution can be rewritten as:
\begin{align*}
    \mathbb{E}[Y^0(t)|A] &= \mathbb{E}\left\{\mathbb{E}[Y^0(t)|X,A]|A\right\} \nonumber \\
    &= \mathbb{E}\left\{\mathbb{E}[Y^0(t)|X]|A\right\} \\
    &= \mathbb{E}[\beta X + \gamma t |A] \\
    &= \beta \mathbb{E}[X|A] + \gamma t
\end{align*}
The second line is true because we assume $Y^0(t) \indep A | X$. Plugging this into the expression for $D(t)$ we can see, 
\begin{equation*}
    D(t) = \mathbb{E}[Y^1(t) - Y^0(t)|A=1] + \beta (\mathbb{E}[X|A=1] - \mathbb{E}[X|A=0])
\end{equation*}
\end{proof}

\end{document}